\documentclass[apl,reprint,a4paper,superscriptaddress,citeautoscript,floatfix,nofootinbib]{revtex4-1}

\pdfoutput=1
\usepackage[charter]{mathdesign}
\usepackage{amsmath}
\usepackage{siunitx}
\usepackage[pdftex]{graphicx}
\usepackage{microtype}
\usepackage{bm}
\usepackage{siunitx}
\usepackage[version=3]{mhchem}

\usepackage[svgnames]{xcolor}

\usepackage[
	colorlinks=True,linkcolor=DarkRed,citecolor=ForestGreen,urlcolor=MediumBlue,
	pdfstartview=FitH,bookmarks=False,pdfpagemode=UseNone
]{hyperref}

\newcommand{\FIG}[1]{Fig.~\ref{#1}}

\newcommand{\Tc}{T_\textrm{c}}
\newcommand{\Jc}{J_\textrm{c}}
\newcommand{\Vg}{V_\textrm{g}}
\newcommand{\sgmatd}{\sigma_{\textrm{3D}}}

\begin{document}

\title{Ionic liquid gating of ultra-thin \ce{YBa2Cu3O_{7-x}} films}

\author{A. F{\^e}te}
\affiliation{Department of Quantum Matter Physics,
Universit{\'e} de Gen{\`e}ve, 24 Quai Ernest-Ansermet, 1211 Gen{\`e}ve 4, Switzerland}
\author{L. Rossi}
\affiliation{Department of Quantum Matter Physics,
Universit{\'e} de Gen{\`e}ve, 24 Quai Ernest-Ansermet, 1211 Gen{\`e}ve 4, Switzerland}
\author{A. Augieri}
\affiliation{Department of Quantum Matter Physics,
Universit{\'e} de Gen{\`e}ve, 24 Quai Ernest-Ansermet, 1211 Gen{\`e}ve 4, Switzerland}
\affiliation{ENEA, Frascati Research Centre, Via E. Fermi, 45, 00044 Frascati, Italy}
\author{C. Senatore}
\affiliation{Department of Quantum Matter Physics,
Universit{\'e} de Gen{\`e}ve, 24 Quai Ernest-Ansermet, 1211 Gen{\`e}ve 4, Switzerland}

\begin{abstract}
In this paper, we present a detailed investigation of the self-field transport properties of an ionic liquid gated ultra-thin \ce{YBa2Cu3O_{7-x}} film. From the high temperature dynamic of the resistivity ($> \SI{220}{\kelvin}$) different scenarios pertaining to the interaction between the liquid and the thin film are proposed. From the low temperature evolution of $\Jc$ and $\Tc$ a comparison between the behavior of our system and the standard properties of YBCO is drawn.
\end{abstract}
\maketitle

Electric field effect doping of superconductors (SC) is a very exciting area of research both from a fundamental and a practical perspective. Interestingly enough, when compared with metals, high temperature superconductors (HTSC) have a relatively low carrier density ($\approx \SI{1e21}{\per \cubic \centi \meter}$) and hence it was quickly proposed that they might be responsive to electrostatic doping. In the 90's, this idea raised the hopes of building high-$\Tc$ transistors \cite{Mannhart1996}. Unfortunately, the solid-state field effect device architectures available at the time realized only modest tuning of the transition temperature ($\Tc$). More recently, due to the large charge accumulation ($\Delta n_\textrm{2D}\approx\SI{e14}-\SI{e15}{\per \square \centi \meter}$) realized using electrical double layer (EDL) transistor structures \cite{Panzer2005,Ono2008}, there was a renew of interest in field effect doping of HTSC and several cuprates, among them \ce{YBa2Cu3O_{7-x}} (YBCO), quite rapidly showed strong tuning of $\Tc$ \cite{Dhoot2010,Bollinger2011,Nojima2011,Leng2011,Leng2012}.

In this work, we carefully study the evolution of $\Tc$ and of the critical current density ($\Jc$) during an EDL-field effect experiment in the underdoped state of YBCO using an ionic liquid (IL). By comparing our low-temperature  data with the literature on underdoped films and bulks, we show that upon negative bias increase of $\Vg$, the transport properties of the device stay coherent with the commonly observed self-field transport properties of doped YBCO. At higher temperatures, by investigating the role and the degree of the interaction between our films and the IL, we identify key parameters and criticalities for EDL gating experiments on YBCO.

4-\SI{5}{\nano \meter} thick YBCO films were deposited on a \ce{TiO_2} terminated \ce{SrTiO_3} (STO) substrate by pulsed laser interval deposition (PLiD) \cite{Koster1999}. A 15 unit cells (uc) \ce{PrBa2Cu3O_{7-x}} (PBCO) buffer layer was used to minimize as much as possible the strain imposed by the substrate. We choosed this modified version of the standard pulsed laser deposition scheme as it leads to a reduced surface roughness (Ra$\approx\SI{1}{\nano \meter}$ by atomic force microscopy) and a higher $\Tc$ ($\approx \SI{70}{\kelvin}$ for capped samples). The KrF excimer laser was set to an \textit{in situ} fluence of \SI{1.3}{\joule \per \square \centi \meter} and to a repetition rate of \SI{10}{\hertz}. The target substrate distance was \SI{53}{\milli \meter}. During growth, the temperature of the substrate was \SI{820}{\celsius} and \SI{0.4}{\milli \bar} of \ce{O2} were inserted in the chamber. Before being cooled to room temperature in about \SI{1}{\hour}, the samples were kept \SI{1}{\hour} at \SI{550}{\celsius} in \SI{0.65}{\bar} of \ce{O2} (\textit{in situ} annealing).

With the aim of defining conducting and insulating areas (Hall bars shapes) on the surface of the substrate, a sacrificial amorphous $\approx \SI{45}{\nano \meter}$ thick \ce{AlO_x} layer (am\ce{AlO_x}) was deposited, at room temperature and prior to the YBCO/PBCO bilayer, on selected areas. During this process, the channel/contacts regions were protected by photoresist. After lift-off, the cuprate layers were grown at high temperature and the gold contacts were evaporated through a shadow mask (\textit{ex situ}). This scheme allowed us to reduce as much as possible the number of post-processing steps after the growth of YBCO.

We used Diethylmethyl(2-methoxyethyl)ammonium bis(trifluoromethylsulfonyl)imide (DEME-TFSI : CAS No. 464927-84-2) as ionic liquid, similarly to \cite{Leng2011}. Care was taken to reduce as much as possible the time spent by the film at room temperature with the IL on it, yet 30 min where necessary for the loading of the device on the probe and its cooldown to \SI{240}{\kelvin}.  This together with YBCO degradation in lab atmosphere \cite{Regier1999} and residual strain from the substrate \cite{Salluzzo2005} is probably responsible for the low $\Tc$ ($\approx\SI{30}{\kelvin}$) realized by the devices prior to any application of gate voltage. The field effect experiments were performed around \SI{240}{\kelvin} were most of the chemical reactions are suppressed but ionic mobility is still substantial \cite{Leng2011}.

Throughout this study only negative voltages were applied to the liquid since this polarity is the one which is injecting holes in YBCO, hence supposed to increase $\Tc$. Secondly, positive gate voltages are strongly suspected to alter oxides by removing oxygen from their structure \cite{Jeong2013,Scherwitzl2010}. During our experiments the leakage current was well below \SI{1}{\nano \ampere}, $\Jc$ was evaluated in self-field at \SI{1}{\micro \volt \per \cm} (fit driven extrapolation) and $\Tc$ at $R=0$.


\begin{figure}[!t]
\centering\includegraphics[width=0.95\columnwidth]{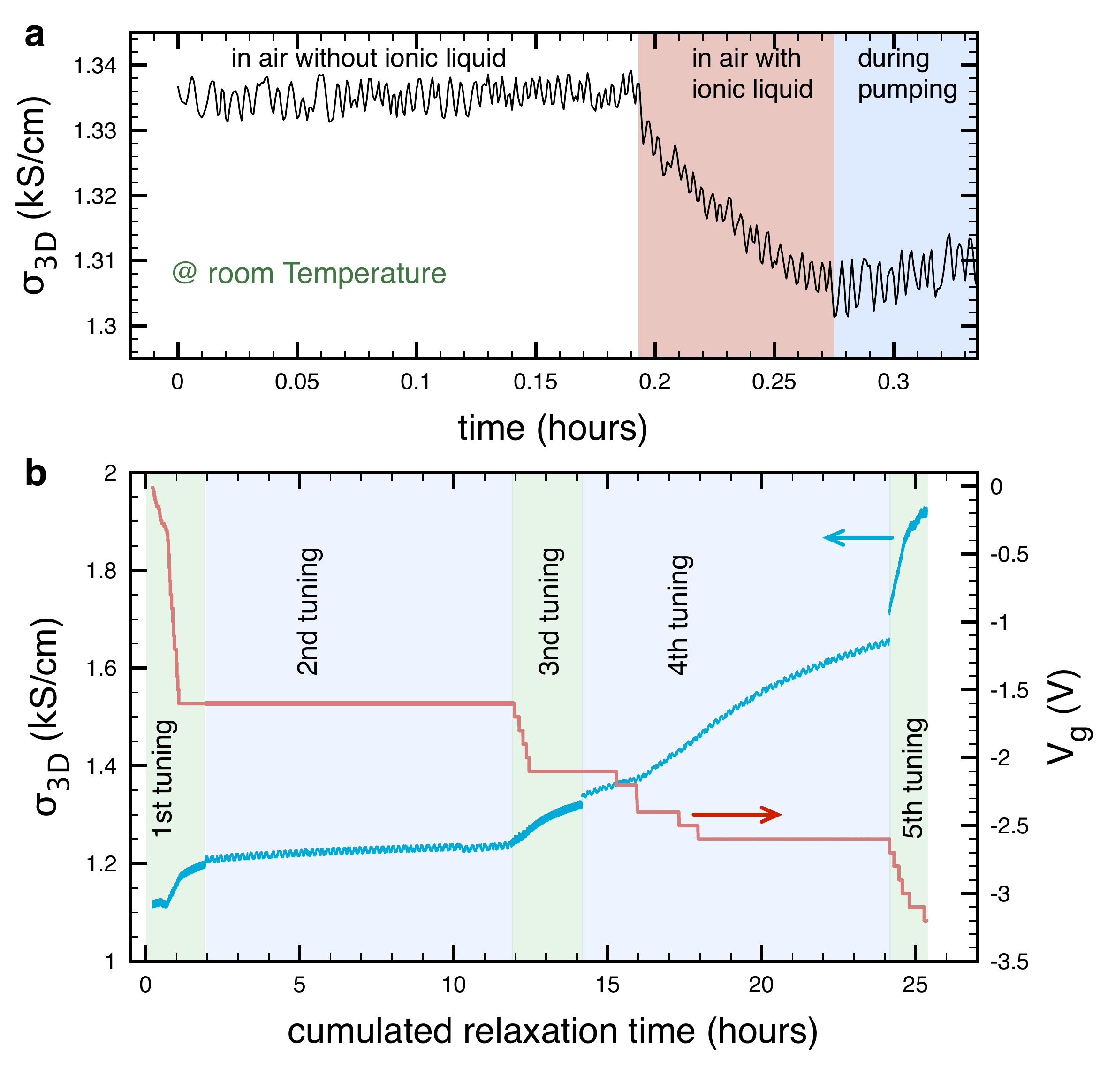}
\caption{\label{Fig2} Evolution of the high temperature conductivity of the device as a function of time. \textbf{a} At room temperature during the application of DEME-TFSI on the channel. \textbf{b} During the field effect experiments (blue curve). Gate voltage is shown in red (right scale). The blue and green areas define the relaxation periods.}
\end{figure}

\FIG{Fig2}a shows the conductivity ($\sgmatd$) of the system while putting it in contact with DEME-TFSI. At this stage the channel is grounded and the liquid is kept floating. First, one sees that the resistance in air is quite stable. The most probable explanation of this observation is that, by corroding, the topmost layers of YBCO protect the deeper layers, which slows down the humidity driven etching of the structure. When the IL is put in contact with the channel we observe a rapid decrease in conductivity, similar to what has been reported in \cite{Leng2011}. As mentioned above, to minimize the degradation of the sample, the measurement probe is then rapidly pumped and cooled down. Surprisingly, \FIG{Fig2}a shows that as soon as the pumping is started conductivity stops decreasing \cite{footnote1}. This observation shows that the decrease in conductivity upon application of the IL is probably not due to uncontrolled electrostatic or temperature changes, as one would hardly understand why pumping would stop one process or the other. Rather, it is generally believed that a low pressure is able to remove water from the ILs. In our case, this would mean that the damages in the thin film are mainly produced by the water contained in the liquid.

Interestingly, it turns out that ILs containing the TFSI anion, despite being hydrophobic, are hygroscopic. In practice, this means that when purchased, the liquid already contains a substantial amount of water. For example, Sigma-Aldrich sells DEME-TFSI with water being the main impurity, at a level of \SI{40}{\milli \mole \per \liter} ($\le$ 500 ppm). In comparison, during our experiments we calculated that water was present in air at a level of only \SI{0.7}{\milli \mole \per \liter} \cite{footnote2}. Finally, we note that DEME-TFSI was recently pointed as a promising liquid for \ce{CO_2} capture \cite{Nonthanasin2014} the latter being probably involved in YBCO decomposition too \cite{Yan1987}.

\begin{figure}
\centering\includegraphics[width=0.85\columnwidth]{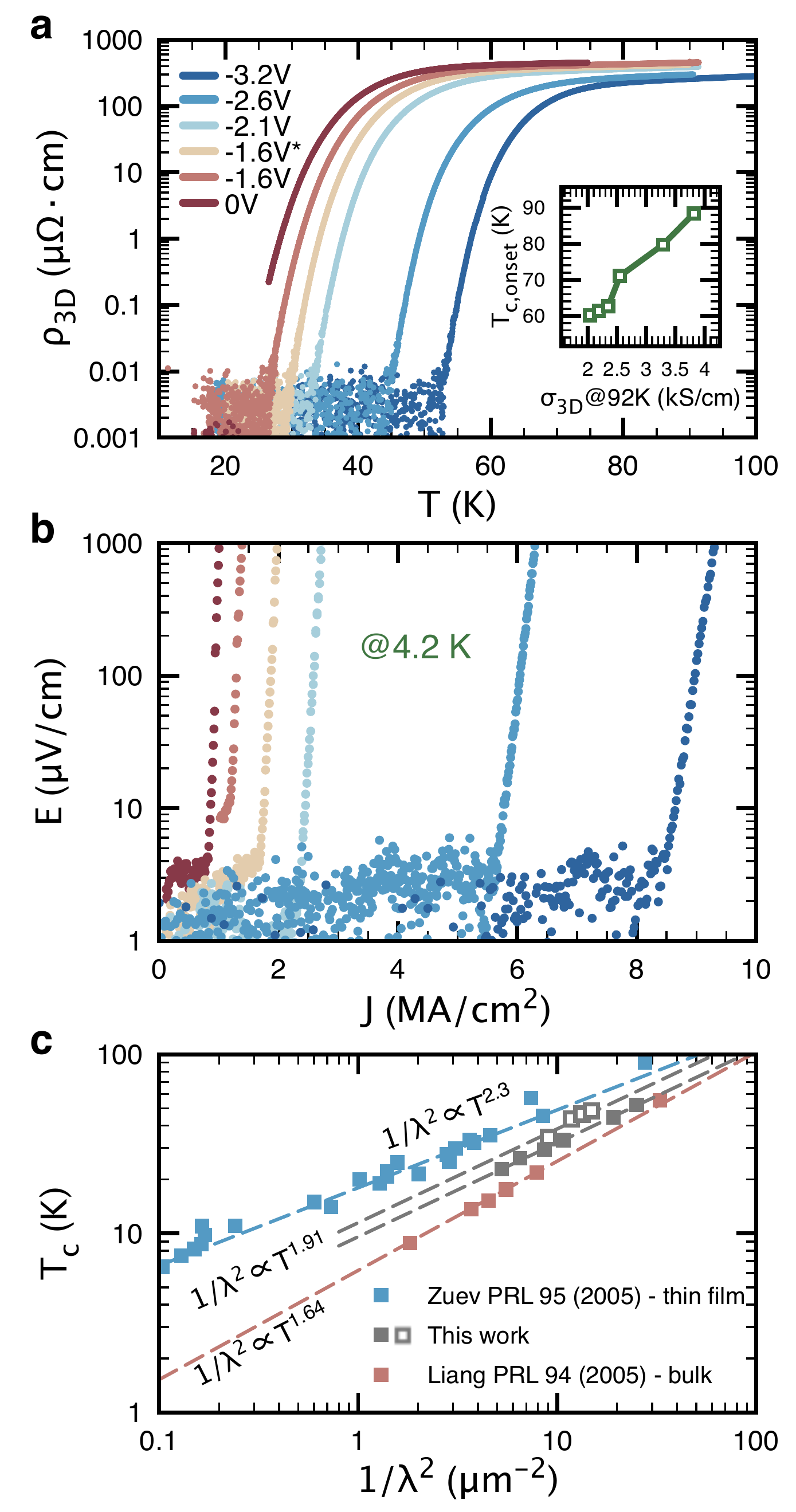}
\caption{\label{Fig3} Low temperature transport properties of the device as a function of field effect. The color code for \textbf{a} and \textbf{b} is the same. \textbf{a} Resistivity as a function of temperature for different doping levels. (inset) Evolution of $T_\textrm{c, onset}$ as a function of conductivity. \textbf{b} Current/voltage characteristics of the device in self-field and at \SI{4.2}{\kelvin}. \textbf{c} Uemura plot showing good agreement between our data and the literature. The fits are shown as dashed lines. The open squares pertain to a second sample measured independently.}
\end{figure}

\FIG{Fig2}b shows the details of the subsequent field effect doping experiments as a function of the cumulated relaxation time. Cumulated relaxation time stands for the fact that in-between the periods of tuning at high temperature (light green and blue areas in \FIG{Fig2}b), temperature is decreased to \SI{4.2}{\kelvin} where current/voltage characteristics (IV) are recorded (see \FIG{Fig3}). Then temperature is increased again and another period of tuning starts.

Strikingly, the tuning of $\sgmatd$ is extremely slow. Indeed, \FIG{Fig2}b clearly shows that upon modification of the gate voltage several hours are necessary to stabilize $\sgmatd$. In fact, $\sgmatd$ almost never stops relaxing completely but since its drift rate is temperature dependent, the resistance of the system is stable in the 4.2-\SI{100}{\kelvin} range \cite{footnote3}.
We note that since during our experiments $I_\textrm{leak}$ is very small and the tuning sessions always increase $\Tc$ (see below), one hardly sees how the electrochemical or intercalation scenario could be at play here.


In \FIG{Fig3}a and b we present the superconducting properties of the EDL field effect device. The inset shows the evolution of $T_\textrm{c, onset}$, taken at \SI{99}{\percent} of the transition. $\sgmatd$ at \SI{92}{\kelvin} is used as an indication of doping. This is a more convenient scale than $\Vg$ due to the relaxation mechanism discussed in \FIG{Fig2} and because it scales with doping, assuming that the normal state mobility of the electrons is not varying significantly. In fact, the legend clearly shows that $\Vg$ cannot be used as a scale since two measurements with the same gate voltage (\SI{-1.6}{\volt} and \SI{-1.6}{\volt}*) but different relaxation times display different $\Tc$ and $\Jc$.

Clearly, EDL gating is very efficient. Changing $\Vg$ from \SI{0}{\volt} to \SI{-3.2}{\volt}, $\Tc(R=0)$ and $\sgmatd$ are increased by $\approx \SI{100}{\percent}$ and $\Jc$ by $\approx \SI{800}{\percent}$. Interestingly, the behavior of $T_\textrm{c, onset}$ as a function of $\sgmatd$ shows a depression which is reminiscent of the one observed in bulk YBCO around \SI{60}{\kelvin} and 1/8 doping. Also, the conjoint shift of $\Tc(R=0)$ and $T_\textrm{c, onset}$ suggests that even in a weak-linked film scenario, the EDL gating is acting homogeneously.

Considering the evolution of $T_\textrm{c, onset}$ as an indication of the doping level of the YBCO film, we calculated the implied variation in the number of hole per copper oxygen plane $\Delta p$. We used the usual  parametrization of $\Tc$ in cuprates  $\Tc/T_{\textrm{c,max}}=1-82.6 (p-0.16)^2$ \cite{Presland1991} (with $T_{\textrm{c,max}}=\SI{93}{\kelvin}$). Yet, in YBCO the \ce{CuO} chains are also contributing to the density of states (DOS) at the Fermi level. To get an estimation of their contribution we relied on \cite{Jarlborg2000} where a ratio of 1.35 was calculated between the DOS at the Fermi level of optimally doped YBCO and \ce{Bi_2Sr_2CaCu_2O_{8+$\delta$}} (the latter being a cuprate without \ce{CuO} chains). Using these approximations, we calculated that during the field effect experiment, each uc of YBCO sees an increase in $\Tc$ corresponding to a carrier density variation of $1.35 \times 2 \times \Delta p \approx  \SI{7e13}{\per \square \square \centi \meter}$. Hence, depending on the actual number of uc of YBCO that survived the damages induced by the transfer from the deposition system to the low temperature one, we infer a total charge density modulation in the few \SI{1e14}{\per \square \square \centi \meter} range. We note that this compares well with the doping induced in other systems using ILs \cite{Fujimoto2013}.

In an effort to better define the link between our measurements and the standard transport properties of YBCO, we made use of the recent work by Talantsev and Tallon which links the self field critical current density of a broad spectrum of superconductors in thin film form  with their London penetration depth $\lambda$ \cite{Talantsev2015}. According to them, experimental data show that as long as the film is thinner than its penetration depth (in YBCO $\lambda > \SI{100}{\nano \meter}$), the self-field $\Jc$ is simply $H_\textrm{c,1}/\lambda$. Making use of the standard Ginzburg-Landau expression for $H_\textrm{c,1}$ leads to $\Jc(\textrm{sf}) \propto \lambda^{-3}$ \cite{footnote4}.

Using this prescription to determine $\lambda$, we present, in \FIG{Fig3}c, our data in the form of a Uemura plot \cite{Uemura1989}. We added results obtained on thin films and bulks \cite{Zuev2005,Liang2005,footnote5}. In the light of the uncertainties on the actual superconducting thickness of our films and hence on the precise value of $\Jc$, the study of the power law relation between $\Tc$ and $\lambda$ is a convenient workaround. Moreover, it is related to the mechanism responsible of the reduction of $\Tc$ in the underdoped state which is a fundamental property of YBCO. As can be observed, the behavior of our device is close from the one of bulk YBCO and in the range of error of \cite{Zuev2005}. Hence, from this point of view, our data are coherent with the transport properties of YBCO in self-field. In the same graph, we also present data obtained from a second sample (open squares) which displays a similar power law evolution. In our opinion, its apparently reduced superfluid density as well as its smaller tunability indicate that the details of the preparation process are key for the performance of the device. Indeed, uncontrolled parameters like lab atmosphere, precise timing for \textit{ex situ} operations and pumping, are probably impacting the interface between the IL and the film.

This result is quite surprising for two reasons. First, it would have been reasonable to think that serious damages to the structure are present at the beginning of the experiment, given the well documented chemical reactivity of YBCO even at room temperature. Secondly, given the ultra-short Thomas-Fermi (TF) screening length of YBCO ($\lambda_\textbf{TF}\approx \SI{1}{\nano \meter}$ \cite{Mannhart1996}), one would have expected that the electrostatic doping is confined to the surface of the thin film. Since $\Tc$ and $\Jc$ are measured with drain-source currents differing by orders of magnitudes, this situation would mean that the probed film thickness depends on the type of measurement. For both reasons, one would not expect a good match between the properties observed here and the ones of bulk samples for example.


\begin{figure}[!t]
\centering\includegraphics[width=0.92\columnwidth]{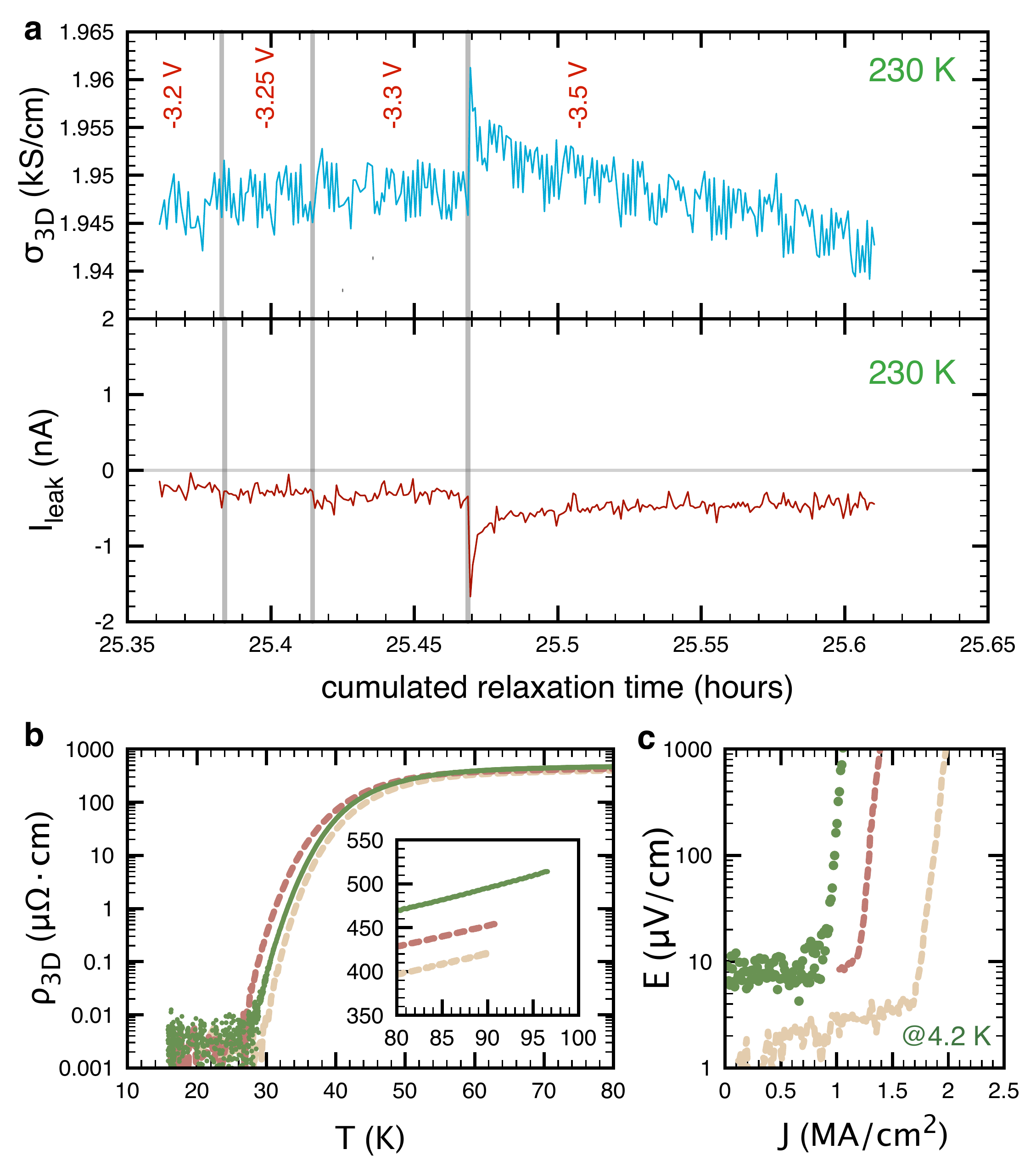}
\caption{\label{Fig4}  \textbf{a} (top) Evolution of the conductivity as a function of time and careful increase towards highly negative gate voltages. (bottom) Leakage current recorded during the same experiment. \textbf{b} Comparison between resistivity versus temperature curves showing similar $\Tc$ but recorded during the initial sweep to negative $\Vg$ (dashed curves) and after the tuning shown in \textbf{a} (green points). The inset is a close-up at higher temperatures. \textbf{c} Current/voltage characteristics at \SI{4.2}{\kelvin} in self-field. The same color code than in \textbf{b} applies.}
\end{figure}

Further experiments are of course needed but, in our opinion, two scenarios can be proposed to explain the apparently homogenous doping of the structure. The first one points that the very high electric field produced by the EDL and the correlations in YBCO are out of the domain of applicability of the TF approximation. Hence the decay of the field in our structure may happen on a larger scale than $\lambda_\textbf{TF}$. We note that, recently, a similar argument was raised by \cite{Piatti2016} to  account for their ability to reduce $\Tc$ in \SI{40}{\nano \meter} thick \ce{NbN} films. The second scenario involves an electric field-driven redistribution of the oxygen atoms in the basal planes of YBCO, as proposed already in the 90's \cite{Chandrasekhar1993}. This possibility was discarded for solid-state field effect devices due their fast response \cite{Schneider1995}, yet, in our case, the huge electric field applied by the EDL and the slow response time of our device are consistent with \cite{Chandrasekhar1993}.

We now present experiments performed right after the 5th tuning at high temperature (see \FIG{Fig2}). As shown in \FIG{Fig4}a, applying a $\Vg$ higher than \SI{-3.2}{\volt} does not lead anymore to an increase of the conductivity. $\sgmatd$ is seen to first saturates (\SI{-3.3}{\volt}) and then starts to decrease with time. In \cite{Leng2012}, this behavior was attributed to the transition of the system to the overdoped state, since $\Tc$ was seen to decrease concomitantly.

What we observed is that this region of $\Vg$ is relatively unstable from the point of view of the leakage current. This is the reason why in \FIG{Fig4}a the data are recorded at \SI{230}{\kelvin} (since $I_\textrm{leak}$ reduces with temperature). Nevertheless, an increase in the absolute value of $I_\textrm{leak}$ can be observed upon increasing $\Vg$. As this behavior is in general attributed to the onset of electrochemical reactions, we decided to not push the experiment to higher values of $\Vg$. Instead, we present in \FIG{Fig4}b and c a comparison between the transport properties of the device before (dashed curves, selected from \FIG{Fig3}) and after (green points) the application of the large negative $\Vg$ of \FIG{Fig4}a \cite{footnote6}. Focusing on the latter, it is apparent that while it displays a similar $\Tc$ than the measurements shown for comparison, its normal state resistivity is increased (similarly to \cite{Leng2012}) and its critical current is reduced by $\approx \SI{40}{\percent}$.

Since we rapidly reduced $\Vg$ when we observed the decrease in conductivity of \FIG{Fig4}a, it must be that the film has been principally damaged during the reduction of the gate voltage. One  possibility is that during the charging process, ions in the liquid bind to the YBCO surface. When the potential is reduced and the ions leave the surface, they take with them some of the atoms of the film. This geometrical process would artificially reduce and increase the estimations of $\Jc$ and $\rho_\textrm{3D}$, respectively, via a reduction of the film thickness. Interestingly enough, there is a large spectrum of molecules which are known to bind chemically with YBCO \cite{Xu1998}, many of them being amines. We note that the DEME ion is a quaternary amine. Moreover, even if here the proposed process is not electrochemical, it was shown recently that etching with an IL can proceed in a controlled way \cite{Shiogai2015}. Further experiments should be performed to elucidate these points but from atomic force microscopy (AFM) performed after EDL gating, we can already say that no topographical damage could be seen down to AFM in-plane resolution ($\approx \SI{10}{\nano \meter}$).

In conclusion, by following the crucial parameters of an EDL gating experiment on YBCO, we have been able to gather substantial information on the evolution of the system while doping. In particular, we have shown that the transport properties of the ultra-thin films are inline with the ones of bulk YBCO. Moreover, we have pointed out water absorption by the IL, field effect driven oxygen motion in the basal plane of YBCO and chemical absorption of molecules on the surface of the thin film as being parameters which might be of primer importance for this type of gating.

The authors would like to thank D.~Zurmuehle and X.~Ravinet for their technical assistance as well as I. Guti\'errez Lezama and C. Berthod for illuminating discussions. Financial support was provided by the SNSF (Grant No. PP00P2\_144673).


\begin{thebibliography}{33}
\bibitem [{Note1()}]{footnote1}%
  \BibitemOpen
  \bibinfo {note} {The small recovery observed upon
  pumping is not commented here and will be the focus of future studies.}\BibitemShut {Stop}%
\bibitem [{Note2()}]{footnote2}%
  \BibitemOpen
  \bibinfo {note} {We also performed gravimetric
  measurements of water absorption by the liquid, which showed that, in lab
  atmosphere, $\approx \SI{40}{\micro \liter}$ of DEME-TFSI absorb water to a
  level of $\approx 1-\SI{2}{\milli \gram}$ on a 5 min timescale. This
  would represents an extra factor 30 in the water content of the IL.}\BibitemShut {Stop}%
\bibitem [{Note3()}]{footnote3}%
  \BibitemOpen
  \bibinfo {note} {This is most probably due to the
  freezing of the motion of the ions below 180 K.}\BibitemShut {Stop}%
\bibitem [{Note6()}]{footnote6}%
  \BibitemOpen
  \bibinfo {note} {The gate voltage at which the
  measurement is performed after the application of large negative $\Vg$ is
  \SI{-2}{\volt}.}\BibitemShut {Stop}%
\end{thebibliography}
\end{document}